\documentclass[aps,pra,reprint,amsmath,amssymb]{revtex4-2}
\usepackage{bm}
\usepackage{graphicx}
\usepackage{mathrsfs}
\usepackage{times}
\usepackage{braket}
\usepackage{xcolor}
\usepackage{mathtools}
\usepackage[
    colorlinks=true,
    linkcolor=blue,
    citecolor=blue,
    urlcolor=blue
]{hyperref}
\usepackage{cleveref}
\crefname{section}{Sec.}{Secs.}
\Crefname{section}{Section}{Sections}
\crefname{equation}{Eq.}{Eqs.}
\Crefname{equation}{Equation}{Equations}
\crefname{figure}{Fig.}{Figs.}
\Crefname{figure}{Figure}{Figures}
\crefname{appsec}{Appendix}{Appendices}
\Crefname{appsec}{Appendix}{Appendices}

\begin{document}
\title{Asymptotics for Frequency Redundancies in Quantum Machine Learning Models}
\author{Felix Paul}
\author{Bhilahari Jeevanesan}
\author{Peter Jung}
\affiliation{Institute of Space Research\\
  German Aerospace Center (DLR)\\
  Berlin, Germany}

\begin{abstract}
  The redundancy distribution of the frequency spectrum has been shown in the literature to impact the expressivity and trainability of Quantum Fourier Models (QFMs).
  In this work, we address the question of how this redundancy spectrum is shaped by the choice of eigenvalues of the data re-uploading Hamiltonians, using a simple mathematical formalism based on generating functions.
  We derive exact and asymptotic redundancy profiles for several structured eigenvalue choices identical for every layer, including arithmetic progressions and single-qubit Pauli encodings, and show that both approach a Gaussian profile as the number of layers grows.
  We then show that this Gaussian limit is not specific to these constructions but is a generic feature of QFMs built from integer eigenvalues that are identical in every layer. These results can be tied to the central limit theorem for random walks.
  These results clarify why generic or unstructured encoding choices give rise to a redundancy bias that favours low frequencies, highlighting the importance of well-thought-out encoding strategies when constructing a QFM.

\end{abstract}

\maketitle

\section{Introduction}
\label{sec:introduction}
Parametrized quantum circuits (PQCs) were first studied from a Fourier analysis standpoint by Vidal and Theis in \cite{vidal2018calculus} and later as Quantum Fourier Models (QFMs) by Schuld et al. in \cite{schuld_fourier_framework}.
It was shown that when the classical input is encoded via the time evolution of a Hamiltonian, the expectation value of a PQC can be represented as a finite trigonometric polynomial in the data variable.
The accessible frequency spectrum $\Omega$ is determined by the differences of the encoding Hamiltonian's eigenvalues, but circuit-dependent constraints can remove theoretically accessible frequencies from the actually available spectrum \cite{qfm_trig_polynomial}.
Repeated data re-uploading \cite{data_reuploading}, which refers to a circuit scheme that consists of alternating data-encoding and trainable blocks, was shown to enlarge the size of the spectrum combinatorially.
Then even shallow or single-qubit circuits can access exponentially many frequencies as the number of re-uploading steps $L$ grows.
In the literature this  scheme is discussed as a central mechanism for controlling the expressivity of QFMs \cite{data_reuploading,schuld_fourier_framework}.

Within this picture, each accessible frequency $\omega$ of the spectrum is generated from different combinations of eigenvalues.
The resulting \emph{frequency redundancy} $r_\omega$, i.e. the number of ways a given frequency $\omega$ can be constructed, has proven to be an important notion.
For example, it was shown that Fourier coefficients associated with low-redundancy frequencies can vanish exponentially with the number of qubits, resulting in a redundancy-driven bias on the expressivity of a QFM~\cite{frequency_redundancy}.
In the same work it was shown, under the assumption that the variational blocks independently form 2-designs, that the variance of the Fourier coefficients is directly controlled by the shape of $r_\omega$.
Additionally, under the same assumptions for the variational blocks, it was shown that the average Fourier \emph{amplitude} spectrum tends towards a Gaussian profile as the number of layers grows~\cite{frequency_profiles_gaussian}.
More generally, the overall spectral power, given by the sum of squared Fourier coefficients, was shown to be exponentially suppressed with the number of qubits under barren plateau conditions~\cite{exp_suppressed_fourier_coeffs}.

Complementing this expressivity picture, a link between redundancy and trainability was established, finding that the magnitude of a coefficient's gradient correlates with its redundancy~\cite{spectral_bias}.
In related work, we found further empirical evidence that frequencies with higher redundancies are associated with more favourable training dynamics~\cite{felix_qfm_trainability}.
This motivates treating $r_\omega$ not just as a spectral artifact but as a property one may wish to actively control.

Given this central role of $r_\omega$ for the analysis of a QFM's capabilities, in this work we deal with the question of how the redundancy spectrum can be engineered by choosing the eigenvalues of the data re-uploading Hamiltonians. 

\noindent {\bf Our main contributions are:}
\begin{itemize}
    \item We derive an integral representation of the frequency redundancy $r_\omega$ in terms of the generating functions of the encoding Hamiltonians' eigenvalues.
    See \cref{eq:generating_function_multilayer_product,eq:redundancy_integral}.
    
    \item For arithmetically growing eigenvalues, we obtain an exact integral expression for $r_\omega$ and derive its asymptotic behaviour in the limit of large layer numbers $L$. We show that this takes on a Gaussian form.
    See \cref{eq:arithmetic_redundancy_integral,eq:arithmetic_redundancy_asymptotic_result_scaled}.
    
    \item For single-qubit encodings, we derive the corresponding generating function and an integral representation of an analytically accessible redundancy distribution.
    See \cref{eq:single_qubit_generating_function,eq:single_qubit_redundancy_integral}.
    
    \item Our central result is that in the case where identical encoding Hamiltonians are used for every layer, the redundancy profile $r_\omega$ converges to a Gaussian whose width is determined by the variance of the eigenvalue distribution.
    See \cref{eq:redundancy_int_asymptotics}.
\end{itemize}
The remainder of this paper is organised as follows:
\Cref{sec:theoretical_background} introduces the theoretical background on QFMs and establishes the generating-function representation of frequency redundancies.
\Cref{sec:redundancy_distributions} applies this formalism to arithmetic eigenvalue spectra and single-qubit Pauli encodings, before deriving the general large-layer asymptotics for fixed encoding eigenvalues.
Finally, \cref{sec:conclusion} summarises the results and discusses directions for the inverse design of redundancy profiles.

\section{Theoretical Background and Calculation of Redundancy Distribution}
\label{sec:theoretical_background}
We begin this section with a short introduction to the notion of Quantum Fourier models introduced in \cite{schuld_fourier_framework}.
One of the central properties of Quantum Fourier Models is their expressivity.
\Cref{sec:qfm_intro} explains how the encoding blocks in the re-uploading scheme contribute to shaping a QFM's spectrum.
In \cref{sec:generating_functions}, the technique of generating functions is introduced and employed to formulate the inverse task of designing the frequency redundancy $r_\omega$ by choosing the encoding Hamiltonian's eigenvalues.

\subsection{Introduction to Quantum Fourier Models}
\label{sec:qfm_intro}

We consider a quantum circuit with $n$ qubits. The objective function of a parametrized quantum circuit (PQC) is usually expressed as the expectation of some operator in the following form
\begin{equation}
  f(x; \theta) = \braket{0\vert U^\dagger(x;\theta) \mathcal{M}U(x;\theta)\vert 0},
  \label{eq:quantum_model}
\end{equation}
where in general $x\in\mathbb{R}^D$ is a $D$-dimensional feature vector, $\theta$ are trainable parameters, $\mathcal{M}$ is an Hermitian observable and $U(x;\theta)$ is a parametrized unitary operation representing the PQC.

\begin{figure}[!tbp]
    \centering
    \includegraphics[width=.8\linewidth]{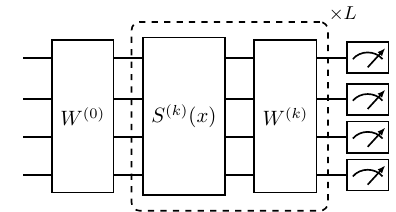}
    \caption{Circuit representation of the re-uploading scheme described in \cref{eq:layered_ansatz}. Encoding and variational blocks are being used in an alternating pattern.}
    \label{fig:layered_circuit}
\end{figure}

Data re-uploading schemes for PQCs often utilize a layered structure according to
\begin{equation}
  U(x;\theta)=\left[\prod_{\ell=1}^L W^{(\ell)}(\theta)S^{(\ell)}(x)\right]W^{(0)}(\theta),
  \label{eq:layered_ansatz}
\end{equation}
where $L$ is the number of layers, $W^{(\ell)}$ are trainable blocks made up of rotational and entangling gates, and $S^{(\ell)}(x)$ are data encoding blocks, where we assume that each feature $x_k$ is encoded as the time-evolution of some Hermitian operator $H_k^{(\ell)}$ in layer $\ell$:
\begin{equation}
  S^{(\ell)}(x)=\prod_{k=1}^D \exp\left(-i x_k H_k^{(\ell)}\right).
  \label{eq:data_encoding_general}
\end{equation}
The individual encoding Hamiltonians from \cref{eq:data_encoding_general} can, without loss of generality, be assumed to be diagonal since the corresponding diagonalizing gates can be absorbed into the trainable blocks.
For simplicity, we assume from hereon that the feature vector $x$ is one-dimensional, i.e. a real number.
We note that properties like the spectrum are straightforward to analyze for higher dimensions see e.g. \cite{schuld_fourier_framework, multidimensional_fourier}.
With both simplifications, the product in \cref{eq:data_encoding_general} collapses to one term and we write the encoding Hamiltonian in layer $\ell$ as
\begin{equation}
  H^{(\ell)}=\text{diag}\big(\lambda_1^{(\ell)},\dots ,\lambda_d^{(\ell)}\big),
  \label{eq:hamiltonian_eigenvalues}
\end{equation}
the $\lambda$'s denoting the eigenvalues and $d=2^n$ being the Hilbert space dimension. We will be interested only in integer eigenvalues, except for a brief paragraph around \cref{eq:arithmetic_redundancy_asymptotic_result_scaled}.

As noted by \cite{schuld_fourier_framework}, we can exploit the diagonal structure of the encoding blocks and write \cref{eq:quantum_model} as a truncated Fourier series in the following form:
\begin{equation}
  f(x;\theta) = \sum_{\omega\in\Omega} c_\omega(\theta)e^{i\omega x},
  \label{eq:quantum_fourier_model}
\end{equation}
where the spectrum $\Omega$ is defined in terms of eigenvalues of the encoding Hamiltonians as
\begin{equation}
  \Omega = \left\lbrace \Lambda_\mathbf{j} - \Lambda_\mathbf{k}\;\Big\vert\; \mathbf{j}, \mathbf{k}\in\left[d\right]^L\right\rbrace \quad \text{with}\quad \Lambda_\mathbf{j} = \sum_{\ell=1}^L \lambda_{j_\ell}^{(\ell)}.
  \label{eq:spectrum_definition}
\end{equation}
Here, $\left[d\right]^L$ denotes the $L$-fold Cartesian product of the set of integers $\left[d\right]\coloneqq \{1,\dots,d\}$ and the  symbol $\mathbf{j}$ denotes a multi-index $\mathbf{j}=\lbrace j_1,\dots j_L\rbrace\in \left[d\right]^L$.

For the analysis, we are interested in the question of how often each frequency occurs in the construction of the spectrum in \cref{eq:spectrum_definition}.
To this end, define the frequency generator set which counts the index pairs $\mathbf{j},\mathbf{k}$ that result in a given frequency
\begin{align}
  R(\omega) & =\left\lbrace(\mathbf{j},\mathbf{k})\in\left[d\right]^L \times[d]^L\; \Big\vert\;\sum_{\ell=1}^L\left(\lambda^{(\ell)}_{j_\ell}-\lambda^{(\ell)}_{k_{\ell}}\right)=\omega\right\rbrace\label{eq:frequency_generator_sum} \\
            & =\left\lbrace(\mathbf{j},\mathbf{k})\in\left[d\right]^L \times[d]^L\; \Big\vert\;\Lambda_\mathbf{j}-\Lambda_\mathbf{k}=\omega\right\rbrace.
  \label{eq:frequency_generator_Lambda}
\end{align}
In the second line we employ the multi-index notation $\Lambda_\mathbf{j}$ for the sum of eigenvalues in \cref{eq:frequency_generator_Lambda}.
The size of the generator set for a given frequency is the central object of interest in this work: The frequency redundancy defined as
\begin{equation}
  r_\omega:= \vert R(\omega)\vert.
  \label{eq:frequency_redundancy_definition}
\end{equation}
In order to employ tools from analysis in the study of $r_\omega$ we next discuss the corresponding generating functions. 

\subsection{Frequency Redundancies from Generating Functions}
\label{sec:generating_functions}

The calculation of $r_\omega$ from \cref{eq:frequency_generator_Lambda} can be elegantly accomplished by using generating functions.  In this section we explain the calculation of $r_\omega$ functions for several examples.

As a simple starting point, let $S$ be a multiset and consider the problem of determining the multiplicities of the elements in the difference multiset $S-S=\lbrace a - b\,\vert\, a,b\in S\rbrace$.
The reason we are considering multisets is that we would like to keep track of multiplicities, e.g. while for a regular set $\{1,1\}=\{1\}$, for a multiset we have $\{1,1\}\neq\{1\}$.

Define the generating function $f_S: \mathbb{C}\to\mathbb{C}$ for the elements of $S$ as 
\begin{equation}
  f_S(z)=\sum_{\lambda\in S} z^\lambda.
  \label{eq:generating_function}
\end{equation}

Since we sum over all elements of $S$, the coefficient of $z^\lambda$ encodes how often the element $\lambda$ occurs in the multiset, i.e. its multiplicity.
Multiplying this function by itself results in a polynomial which encodes the multiplicities of any element of the sum multiset $S+S$ \cite{graham94, additive_combinatorics}:
\begin{equation}
  f_S(z)^2=\left(\sum_{a\in S}z^a \right)\left(\sum_{b\in S}z^b\right)=\sum_{\lambda\in S+S}z^\lambda,
  \label{eq:sumset_example}
\end{equation}
Similarly, by observing the generating function with argument $z$ replaced by $z^{-1}$ yields the same multiset but with the signs of all the elements switched, i.e. $f_S\left(z^{-1}\right)=f_{-S}(z)$, one obtains a similar expression for the difference multiset $S-S$:
\begin{equation}
  f_S(z) f_S\left(z^{-1}\right)=\sum_{\omega\in S-S}z^\omega.
  \label{eq:spectrum_generating_function}
\end{equation}
Now, the QFM's spectrum $\Omega$ is defined in \cref{eq:spectrum_definition}. 
Let us denote the multiset of the encoding Hamiltonian's eigenvalues in layer $\ell$ from \cref{eq:hamiltonian_eigenvalues} by
\begin{eqnarray}{\Lambda^{(\ell)}:=\lbrace\lambda_j^{(\ell)}\;|\;j \in [d]\rbrace}.
\end{eqnarray}
For a single layer, i.e. $L=1$, the frequencies of the model are identical to the difference multiset $\Lambda^{(1)}-\Lambda^{(1)}$. Then we can use the identity \cref{eq:spectrum_generating_function} with $S=\Lambda^{(1)}$ to obtain a formula for the frequency redundancy $r_\omega$ in terms of $f_{\Lambda^{(1)}}(z)$:
\begin{eqnarray}
f_{\Lambda^{(1)}}(z) f_{\Lambda^{(1)}}(z^{-1})  =\sum_{\omega\in \Lambda^{(1)}-\Lambda^{(1)}}z^\omega = \sum_{\omega=-\infty}^{+\infty} r_\omega z^\omega. \label{eq:spectrum_generating_functionL1}
\end{eqnarray}
If the left-hand side is analytically tractable, it may be possible to calculate $r_\omega$ in closed form.

For $L>1$, we can generalize \cref{eq:spectrum_generating_functionL1} to a product over the eigenvalue multisets $\Lambda^{(\ell)}$ of each individual layer according to
\begin{equation}
  f(z) = \prod_{\ell=1}^L f^{(\ell)}(z)\prod_{\ell^\prime=1}^L f^{(\ell^\prime)}(1/z) = \sum_{\omega=-\infty}^{+\infty} r_\omega z^\omega
  \label{eq:generating_function_multilayer_product}
\end{equation}
with
\begin{eqnarray}
  \qquad f^{(\ell)}(z):=\sum_{\lambda\in\Lambda^{(\ell)}}z^\lambda.
  \label{eq:generating_function_multilayer}
\end{eqnarray}

If  one is able to find a closed form for the generating function $f$ in \cref{eq:generating_function_multilayer}, one can recover the frequency redundancy from Cauchy's integral formula 
\begin{equation}
  r_\omega = \frac{1}{2\pi i}\oint_\gamma dz\, \frac{f(z)}{z^{\omega+1}} = \frac{1}{2\pi}\int_{-\pi}^{\pi}d\theta\, f(e^{i\theta}) e^{-i\omega\theta}.
  \label{eq:redundancy_integral}
\end{equation}
by integrating around the unit circle. This is possible because the $\omega's$ are integer-valued. In the second step we parametrized the unit circle by $z=e^{i\theta}$. 
We will use this approach below to calculate exact and asymptotic forms of frequency redundancies.

Starting from the set of eigenvalues of the encoding Hamiltonians, the calculation of the redundancies $r_\omega$ is in principle well-defined.
Much harder than this forward-direction, is the corresponding inverse problem: given a desired redundancy distribution $r_\omega^{\text{target}}$, find an eigenvalue multiset that results in a redundancy profile $r_\omega$ that is close to the desired target.
This problem has been studied in various disguises in different communities and is also related to the famous turnpike problem \cite{skiena1990reconstructing}.
In the QFM setting, this connection has recently been used to investigate when frequency spectra and their multiplicities determine the underlying encoding eigenvalues \cite{spectral_invariance_area_transformation}.

\section{Redundancy Distributions from Generating Functions}
\label{sec:redundancy_distributions}
In this section we first present some illustrative examples of encoding Hamiltonians and calculate their redundancy profiles using the approach of the previous section.
Following this, we show that when choosing an encoding scheme with eigenvalues that do not change from layer to layer, the resulting redundancy distribution $r_\omega$ will asymptotically tend to a Gaussian profile for a large number of layers.

\subsection{Example: Arithmetically Growing Eigenvalues}
\label{sec:arithmetic_progression}
As a first example, consider a setting with identical diagonal Hamiltonians \cref{eq:hamiltonian_eigenvalues} for every layer.
Thus, one has $\lambda_k^{(\ell)}\equiv \lambda_k$, for all $k\in\left[d\right]$.
Moreover, let the eigenvalues be the sequence of integers $\lambda_k = k$, in other words the encoding Hamiltonian is  
\begin{equation}
    H=\text{diag}\left(1, 2, \ldots , d\right).
\end{equation}
Then each factor in the products in \cref{eq:generating_function_multilayer_product} is identical. Thus the generating function $f(z)$ is a geometric series and can be summed in closed form:
\begin{equation}
  f(z)=\left(\sum_{k=1}^d z^k\right)^L\left(\sum_{k=1}^d z^{-k}\right)^L= \frac{1}{z^{L(d-1)}}\left(\frac{z^d-1}{z-1}\right)^{2L}.
  \label{eq:arithmetic_generating_function}
\end{equation}
This function has to be inserted into \cref{eq:redundancy_integral}, thus yielding the following integral expression for the redundancy
\begin{equation}
  r_\omega = \frac{1}{2\pi}\int_{-\pi}^{\pi} d\theta\, \left(\frac{\sin(d\theta/2)}{\sin(\theta/2)}\right)^{2L}\cos(\omega\theta),
  \label{eq:arithmetic_redundancy_integral}
\end{equation}
where the term in parentheses is the Dirichlet kernel.
For growing $d$ it is concentrated around $x=0\equiv 2\pi$.
We use this in \cref{app:arithmetic_progression_integral} in order to derive the large-$L$ asymptotic form of the integral:
\begin{equation}
    r_\omega\sim d^{2L}\sqrt{\frac{3}{\pi L(d^2-1)}}\exp\!\left(-\frac{3\omega^2}{L(d^2-1)}\right)\ \text{for}\ \omega\in\mathbb{Z},
  \label{eq:arithmetic_redundancy_asymptotic_result}
\end{equation}
where $\sim$  denotes asymptotic equivalence as $L\to\infty$.
Thus, when the eigenvalues $\lambda_k$ are the integers $1,\dots,d$, we obtain a redundancy that decays exponentially in $\omega^2$. The width of the Gaussian clearly scales like $\sqrt{L}$.

This immediately generalizes to equidistant eigenvalues of the form
\begin{equation}
    H=\text{diag}(a+c, a+2c, \ldots, a+dc)\quad,a\in \mathbb{R},c\in \mathbb{R}\setminus{\{0\}}
    \label{eq:arithmetic_eigenvalues_scaled}
\end{equation}
To this end, we inspect \cref{eq:spectrum_definition}.
Firstly, the constant shift $a$ cancels out since only differences of $\lambda$'s enter $\Omega$.
Secondly, the factor $c$ scales the possible frequencies in $\Omega$ according to $\omega^\prime = c \cdot\omega$ compared to the case discussed above. Thus all frequencies in $\Omega$ are integer multiples of $c$ and we have
\begin{equation}
    r_{\omega}\sim\begin{dcases}
    \sqrt{\frac{3}{\pi L}}d^{2L-1}\exp\left(-\frac{3\omega^{2}}{Lc^{2}d^{2}}\right) & \text{for}\ \omega\in c\mathbb{Z},\\
0 & \text{otherwise}.
\end{dcases}\
\label{eq:arithmetic_redundancy_asymptotic_result_scaled}
\end{equation}
To summarize, the redundancy formulas for arithmetic progressions have asymptotic forms that are Gaussian in the large $L$ regime.
\begin{figure}[!tbp]
    \centering
    \includegraphics[width=.9\columnwidth]{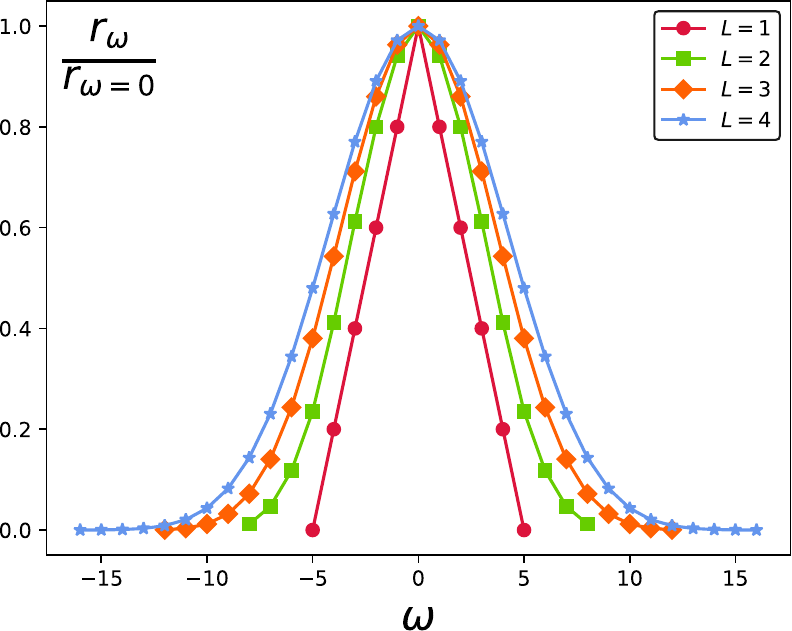}
    \caption{Normalized redundancy profiles for the arithmetic progression of eigenvalues for varying values of $L$ at $d=5$.
    For $L=1$, $r_\omega$ is a triangle function, see \cref{eq:arithmetic_redundancy_integral_L1}.
    As $L$ grows, the profile becomes an increasingly narrow Gaussian.}
    \label{fig:redundancy_asymptotics_arithmetic_L}
\end{figure}

\begin{figure}[!tbp]
    \centering
    \includegraphics[width=.9\columnwidth]{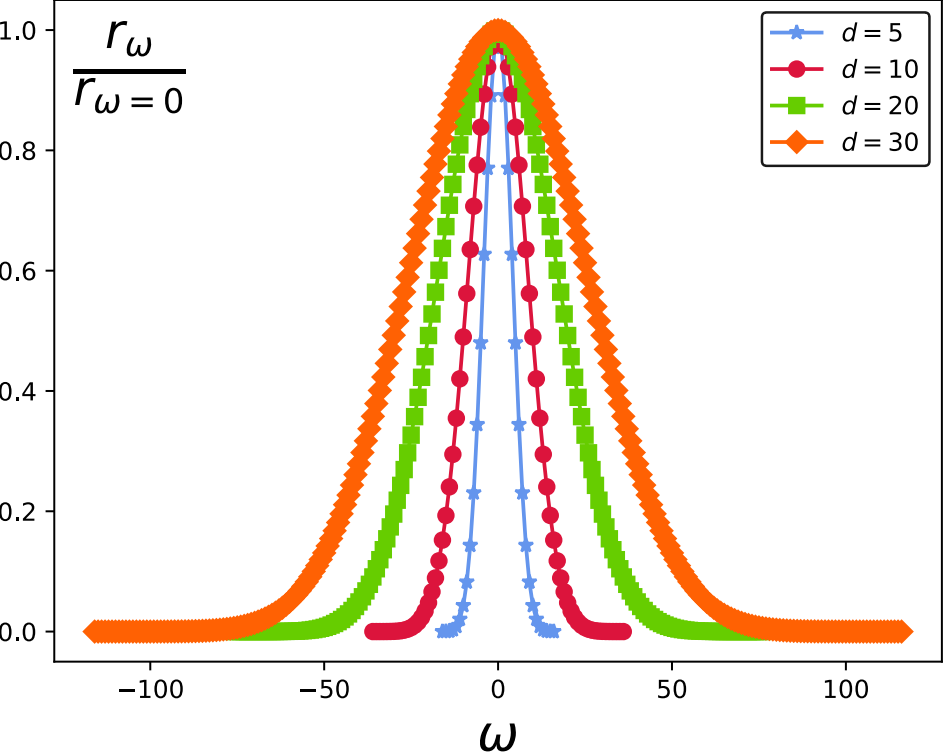}
    \caption{Normalized redundancy profiles for the arithmetic progression of eigenvalues with $L=4$ and varying $d$.}
    \label{fig:redundancy_asymptotics_arithmetic_d}
\end{figure}
It is interesting to note that in the single layer case, i.e.  for $L=1$, the redundancy is a triangle function, see \cref{fig:redundancy_asymptotics_arithmetic_L}.
In fact, we show in Appendix \ref{app:triangle_function} that the case $L=1$ can be treated exactly and yields
\begin{eqnarray}
  r_\omega &=& \frac{1}{2\pi}\int_{-\pi}^{\pi} d\theta\, \left(\frac{\sin(d\theta/2)}{\sin(\theta/2)}\right)^{2}\cos(\omega\theta)\\&=&\begin{cases}
d-|\omega| & \text{for}\ |\omega|\leq d,\\
0 & \text{otherwise}.
\end{cases}
  \label{eq:arithmetic_redundancy_integral_L1}
\end{eqnarray}

\Cref{fig:redundancy_asymptotics_arithmetic_L} illustrates the emerging Gaussian profile for increasing values of $L$ while \cref{fig:redundancy_asymptotics_arithmetic_d} highlights the increasing width of the Gaussian envelope with growing $d$ at fixed $L$.

\subsection{Example: Single-Qubit Gates}
\label{sec:single_qubit_gates}
The last section showed that when the eigenvalues $\lambda_k$ form an arithmetic sequence, a Gaussian form for the redundancy $r_\omega$ is obtained.
This already indicates that the convergence of $r_\omega$ to a Gaussian with increasing layer size $L$ is a generic phenomenon. 
Before we delve into the precise statement of this fact in the next section, let us first consider another calculable example. This one is especially relevant for data encoding in the QML setting.

\begin{figure}[!tbp]
    \centering
    \includegraphics[width=.8\columnwidth]{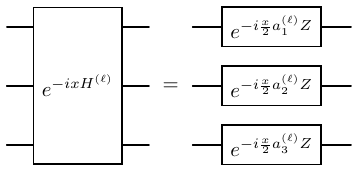}
    \caption{Illustrative circuit representation of \cref{eq:single_qubit_encoding} for three qubits.}
    \label{fig:single_qubit_gates}
\end{figure}

Let the data encoding block $S^{(\ell)}(x)$ of \cref{eq:data_encoding_general} be a product of Pauli-$Z$ single-qubit rotation gates
\begin{equation}
    S^{(\ell)}(x) = \exp\left(-i x H^{(\ell)}\right)=\bigotimes_{k=1}^n \exp\left(-i \frac{x}{2} a_k^{(\ell)}Z\right)
    \label{eq:single_qubit_encoding}
\end{equation}
where $a_k^{(\ell)}\in \mathbb{Z}$ are chosen for each layer $\ell$ (see \cref{fig:single_qubit_gates} for an illustration).
The eigenvalues of $H^{(\ell)}$ are given by the multiset
\begin{equation}
    \Lambda^{(\ell)}=\left\{\frac{1}{2}\sum_{k=1}^{n}\sigma_i a_k^{(\ell)} \,\bigg|\, \sigma_i\in\{\pm 1\}\right\}.
    \label{eq:single_qubit_eigenvalues}
\end{equation}
Then the full spectrum of the QFM is given by
\begin{equation}
    \Omega = \left\{\frac{1}{2}\sum_{\ell=1}^L\sum_{k=1}^{n} a_k^{(\ell)} (\sigma_k- \sigma_k') \;\bigg|\;\sigma_k,\sigma_k'\in\{\pm1\}\right\},
\end{equation}
With the spectrum defined, we can write down the generating Laurent polynomial as
\begin{multline}
    f^{(\ell)}(z) = \sum_{\lambda\in\Lambda^{(\ell)}}z^\lambda = \sum_{\sigma_{1},\dots,\sigma_{n}\in\{\pm1\}}z^{\frac{1}{2}\sum_{k=1}^{n}\sigma_{k}a^{(\ell)}_{k}}\\
    =\prod_{k=1}^{n}\left(z^{\frac{1}{2}a^{(\ell)}_{k}}+z^{-\frac{1}{2}a^{(\ell)}_{k}}\right)=2^n\prod_{k=1}^{n}\cos\left(\frac{1}{2}a^{(\ell)}_{k}\theta\right),
\end{multline}
where in the last step we have set $z=e^{i\theta}$ since the contour integration in \cref{eq:redundancy_integral} proceeds along the unit circle.
From \cref{eq:generating_function_multilayer}, the full generating function is given by
\begin{equation}
f(e^{i\theta}) = 4^{n L}\prod_{\ell=1}^{L}\prod_{k=1}^{n}\cos^2\left(\frac{1}{2}a^{(\ell)}_{k}\theta\right).
\label{eq:single_qubit_generating_function}
\end{equation}
This expression cannot be evaluated in closed form for general $a_k^{(\ell)}$ but there is a particularly simple case that can be evaluated analytically.
It is based on the trigonometric identity
\begin{equation}
 \frac{\sin{(2^n \theta)}}{\sin{\theta}}= 2^n\prod_{k=1}^{n} \cos(2^{k-1} \theta)
\end{equation}
which follows from the duplication formula $\sin(2^m x)=2\sin(2^{m-1}x)\cos(2^{m-1}x)$ by iteratively unfolding the $\sin$ factor on the right hand side using the same duplication formula. 

Thus, for the special case where the coefficients are identical for every layer, i.e. $a_k^{(\ell)}\equiv a_k\;\forall \ell$, and are set to $a_k = 2^{k-1}$ we obtain the closed formula
\begin{equation}
    f(e^{i\theta}) = \left(\frac{\sin{(2^n \theta/2)}}{\sin{(\theta/2)}}\right)^{2L},
\end{equation}
from which the redundancy follows as
\begin{eqnarray}
    r_\omega = \frac{1}{2\pi}\int_{-\pi}^{\pi}d\theta\, \left(\frac{\sin(2^{n}\theta/2)}{\sin(\theta/2)}\right)^{2L}\cos\left(\omega\theta\right).
    \label{eq:single_qubit_redundancy_integral}
\end{eqnarray}
We recognize here again the Dirichlet kernel raised to the $2L$ power, just as in \cref{eq:arithmetic_redundancy_integral} for arithmetically growing eigenvalues.
This somewhat surprising result is explained by the fact that for $a_k=2^{k-1}$ the eigenvalues in \cref{eq:single_qubit_eigenvalues} are exactly the numbers $\{-(d+1)/2,-(d+1)/2+1,\dots,(d+1)/2\}$. This corresponds precisely to the case of equidistant eigenvalues from \cref{eq:arithmetic_eigenvalues_scaled} with shift $a=-(d+1)/2$ and scaling $c=1$ (see also \cite{generalization_despite_overfitting}).

Thus, the redundancy in \cref{eq:single_qubit_redundancy_integral} is once more the triangle function for $L=1$, while for increasing $L$ the shape approaches a Gaussian. 

\subsection{Redundancy for Fixed Eigenvalues}
\label{sec:fixed_eigenvalues}
In order to gain insight into how much freedom and flexibility one has in shaping the redundancy profile, we consider now a scenario where the same encoding Hamiltonian is used in every layer.
Denoting the eigenvalues of the Hamiltonian by $\{\lambda_k\}_{k=1}^d$, we assume that each $\lambda_k \in [M]=\{1,2,\dots,M\}$. 
In this setting, the products of \cref{eq:generating_function_multilayer} collapse to multiples of a single-layer generating function and one can write it in its angle-parametrized form as
\begin{equation}
  f_H(e^{i\theta})=\left(\sum_{k=1}^d e^{i\lambda_k\theta}\right)^L \left(\sum_{k=1}^d e^{-i\lambda_k\theta}\right)^L = d^{2L}\vert h(\theta)\vert^{2L}\,,
  \label{eq:averaging_generating_function}
\end{equation}
with the function $h(\theta)$ defined as
\begin{equation}
    h(\theta)\coloneq\frac{1}{d}\sum_{k=1}^d e^{i\lambda_k \theta}.
    \label{eq:sum_complex_rvs}
\end{equation}
The redundancy can then be calculated via the integral of \cref{eq:redundancy_integral} according to
\begin{eqnarray}
  r_\omega &=& \frac{d^{2L}}{2\pi}\int_{-\pi}^\pi d\theta\, e^{-i\omega\theta} \vert h(\theta)\vert^{2L} \\
  &=& \frac{d^{2L}}{2\pi}\int_{-\pi}^\pi d\theta\, e^{-i\omega\theta} e^{{2L}\log\vert h(\theta)\vert } .
  \label{eq:redundancy_int_arithmetic}
\end{eqnarray}
We are interested in the asymptotic behaviour of \cref{eq:redundancy_int_arithmetic} for a large number of layers $L$.
To this end, we employ Laplace's method~\cite{bender_orszag}, according to which the behavior of the integral for large $L$ is dominated by the behavior of the integrand near the maximum of the exponent. We find that the frequency redundancy can be written asymptotically as
\begin{equation}
    r_\omega\sim \frac{d^{2L}}{\sqrt{4\pi Ls^2}}\exp\!\left(-\frac{\omega^2}{4Ls^2}\right)\,,\qquad L\to\infty\,,
    \label{eq:redundancy_int_asymptotics}
\end{equation}
with 
\begin{eqnarray}
    s^2&=&\frac{1}{d}\sum_{k=1}^d (\lambda_k-\bar{\lambda})^2\\
\bar{\lambda}&=&\frac{1}{d}\sum_{k=1}^d \lambda_k.
\end{eqnarray}
For a detailed derivation see \cref{app:asymptotic_redundancy_integral}.
In the derivation we assume that the eigenvalues differences do not have a greatest common divisor $g>1$. This is not a serious limitation, since we can always shift and divide the eigenvalues by $g$ in order to make $g=1$, see the discussion around \cref{eq:arithmetic_eigenvalues_scaled}. 

\cref{eq:redundancy_int_asymptotics} can be interpreted as a statement that for a generic setting where the eigenvalues of the encoding Hamiltonian do not have much structure, the redundancy of a frequency $\omega$ has a Gaussian suppression in $\omega$.
Intuitively, it is not surprising that redundancy decreases as $|\omega|$ increases: frequencies with large magnitudes can only be achieved by the largest eigenvalues with the same sign.

However, it is not obvious that in the limit of a large number of layers with identical encoding Hamiltonians the asymptotic behavior of the redundancies is a Gaussian distribution, independent of the dimension of the Hilbert space $d$. This is an aspect of universality that can be illuminated by a random walk interpretation. Since $h(\theta)^L$ is the characteristic function of an $L$-step walk with increments drawn i.i.d. from the fixed eigenvalue set, $r_\omega/d^{2L}$ is essentially the probability of such a walk landing at $\omega$ after $L$ steps.
The Gaussian profile therefore reflects the central-limit behaviour expected after many such iterations.

\section{Conclusion \& Outlook}
\label{sec:conclusion}
In this work, we introduce a formalism that allows for the analysis of a QFM's frequency redundancy profile by examining the Fourier transform of a generating function, depending on the encoding Hamiltonians' eigenvalues.
We use this formalism to derive exact and asymptotic profiles for different settings, i.e. arithmetically growing eigenvalues and single-qubit encoding gates, as well as for the case where the encoding Hamiltonian is the same in  every layer.
We find in all these cases that the redundancy converges to a Gaussian profile when the number of layers $L$ is increased. While our study concerns the frequencies of QFMs, we note that there exist complementary results about the Fourier amplitudes of QFMs that showed that the absolute amplitudes approach a Gaussian profile~\cite{frequency_profiles_gaussian}.

Our results strongly suggest that simple or unstructured choices for data-encoding Hamiltonians are unlikely to yield a redundancy profile that treats all frequencies equally. In the literature, the  latter is considered as a prerequisite for having unbiased QFMs since a bias towards frequencies with high redundancies exists when considering gradient magnitudes or the flexibility of the model's Fourier coefficients \cite{frequency_redundancy,spectral_bias,felix_qfm_trainability}.

A natural followup to our work would be to tackle the inverse design problem of realizing target redundancy profiles and to what extent it is possible to arrive at an unbiased QFM.

\subsection*{Acknowledgments}
This project was made possible by the DLR Quantum Computing Initiative and the Federal Ministry for Research, Technology and Space; \href{qci.dlr.de/projects/qcoptsens/}{qci.dlr.de/projects/qcoptsens/}

\newpage

\appendix
\crefalias{section}{appsec}
\onecolumngrid

\begin{center}
  \rule{0.5\linewidth}{0.5pt}
\end{center}

\section{Redundancy Asymptotics for Arithmetically Increasing Eigenvalues}
\label{app:arithmetic_progression_integral}
Here we sketch the asymptotic evaluation of \cref{eq:arithmetic_redundancy_integral} for arithmetically growing eigenvalues $\lambda_k = k\in [d]$, taking $L \rightarrow\infty$ at fixed $d$.
We define the normalized function
\begin{equation}
    h(\theta) := \frac{1}{d}\,\frac{\sin(d\theta/2)}{\sin(\theta/2)},
    \label{eq:app_normalized_h_func}
\end{equation}
so that \cref{eq:arithmetic_redundancy_integral} can be written as
\begin{equation}
    r_\omega = \frac{d^{2L}}{2\pi} \int_{-\pi}^{\pi} d\theta\, h(\theta)^{2L}  \cos(\omega\theta).
    \label{eq:app_arithmetic_redundancy_integral_h}
\end{equation}
We have $h(0) = 1$ and $|h(\theta)| \leq 1$ on $[-\pi,\pi]$, with the maximum at $\theta = 0$ being unique for fixed $d$.
We expand $h(\theta)$ locally around $\theta=0$ and keep only the dominant terms:
\begin{equation}
    h(\theta) = \frac{\frac{d\theta}{2} - \frac{(d\theta/2)^3}{6} + O(\theta^5)}{d\left(\frac{\theta}{2} - \frac{(\theta/2)^3}{6} + O(\theta^5)\right)}
    = \left(1 - \frac{d^2\theta^2}{24} + O(\theta^4)\right)\left(1 + \frac{\theta^2}{24} + O(\theta^4)\right),
\end{equation}
hence
\begin{equation}
    \ln h(\theta) = -\frac{d^2 - 1}{24}\,\theta^2 + O(\theta^4),
\end{equation}
when expanding the logarithm.
This means that
\begin{equation}
    h(\theta)^{2L} = \exp\!\left(-\frac{L(d^2-1)}{12}\,\theta^2 + L\,O(\theta^4)\right).
\end{equation}
The integral in \cref{eq:app_arithmetic_redundancy_integral_h} is dominated by $|\theta| = O(1/\sqrt{L})$, so the higher-order terms in the exponent are suppressed by a factor of order $1/L$ and may be neglected at leading order, which means that
\begin{equation}
    h(\theta)^{2L}\sim \exp\!\left(-\frac{L(d^2-1)}{12}\,\theta^2\right),\qquad L \rightarrow \infty.
    \label{eq:app_arithmetic_h_asymptotic}
\end{equation}
Outside the window $|\theta| = O(1/\sqrt{L})$ the integrand is already exponentially suppressed, since $h(\theta) < 1$ for $\theta \neq 0$ for fixed $d$, so we may extend the integration limits to $\pm\infty$ at asymptotically negligible cost.
Inserting \cref{eq:app_arithmetic_h_asymptotic} into \cref{eq:app_arithmetic_redundancy_integral_h} leads to
\begin{equation}\label{eq:appAIntegral}
    r_\omega \sim \frac{d^{2L}}{2\pi}\int_{-\infty}^{\infty} d\theta\,\exp\!\left(-\frac{L(d^2-1)}{12}\theta^2\right)\cos(\omega\theta),\qquad L\rightarrow\infty.
\end{equation}
For $a>0$ and $b\in\mathbb{R}$ we use the identity
\begin{equation}
    \int_{-\infty}^{\infty} dx\,
    e^{-a x^2} \cos(b x)
    = \sqrt{\frac{\pi}{a}} \exp\!\left(-\frac{b^2}{4a}\right),
\end{equation}
to evaluate the integral \cref{eq:appAIntegral}
\begin{equation}
    r_\omega \sim d^{2L}\sqrt{\frac{3}{\pi L (d^2-1)}}\,\exp\!\left(-\frac{3\omega^2}{L(d^2-1)}\right),
    \qquad L \rightarrow \infty.
\end{equation}

\section{Triangle Function for One Layer}
\label{app:triangle_function}
The redundancy integral for arithmetically growing eigenvalues for $L=1$ is
\begin{equation}
  r_\omega
  = \frac{1}{2\pi}\int_{-\pi}^{\pi}d\theta
  \left(\frac{\sin(d\theta/2)}{\sin(\theta/2)}\right)^2
  \cos(\omega\theta)\,,
  \label{eq:app_redundancy_triangle}
\end{equation}
with $d=2^n,\,n\in\mathbb{N}$ and $\omega\in\mathbb{Z}$.
The integral can be evaluated by using the sum representation of the Fej\'er kernel $F_N(\theta)$ of order $N$~\cite{harmonic_analysis}:
\begin{equation}
  F_N(\theta)
  = \frac{1}{N+1}
  \left(\frac{\sin((N+1)\theta/2)}{\sin(\theta/2)}\right)^2
  = \sum_{k=-N}^{N}\left(1-\frac{|k|}{N+1}\right)e^{ik\theta}\,.
  \label{eq:app_fejer_kernel}
\end{equation}
For $N=d-1$ this gives
\begin{equation}
  \left(\frac{\sin(d\theta/2)}{\sin(\theta/2)}\right)^2
  = \sum_{k=-(d-1)}^{d-1}(d-|k|)e^{ik\theta}.
  \label{eq:app_kernel_fourier_series}
\end{equation}
Now, by orthogonality on we have
\begin{equation}
  \frac{1}{2\pi}\int_{-\pi}^\pi d\theta\,e^{ik\theta}\cos(\omega\theta)=\frac{1}{2}\left(\delta_{k,\omega}+\delta_{k,-\omega}\right)\,.
  \label{eq:app_fourier_orthongonality}
\end{equation}
Combining \cref{eq:app_kernel_fourier_series} and \cref{eq:app_fourier_orthongonality} for the evaluation of the integral results in
\begin{equation}
  r_\omega =
  \begin{dcases}
    d-|\omega|, & |\omega|<d,    \\[1mm]
    0,          & |\omega|\ge d.
  \end{dcases}
  \label{eq:app_redundancy_triangle_result}
\end{equation}

\section{Redundancy Asymptotics for Fixed Eigenvalues}
\label{app:asymptotic_redundancy_integral}
For a fixed set of eigenvalues $\lambda_1,\ldots,\lambda_d \in \left[M\right]$, define $h(\theta) = \frac{1}{d}\sum_{k=1}^d e^{i\theta \lambda_k}$. We wish to calculate the frequency redundancy given by the integral
\begin{equation}
    r_\omega = \frac{d^{2L}}{2\pi}\int_{-\pi}^{\pi} d\theta\,e^{-i\omega\theta}\,|h(\theta)|^{2L}.
\label{eq:app_redundancy_integral}
\end{equation}
This will be computed in the asymptotic limit where the number of layers $L$ is large.
To this end, we employ Laplace's method~\cite{bender_orszag}, for which we have to expand $\vert h(\theta)\vert^{2}$ around it's global maximum on the interval of integration.
Since $\vert h(\theta)\vert \leq \frac{1}{d}\sum_k \vert e^{i\theta \lambda_k}\vert=1$ we find one obvious global maximum at $\theta=0$. 
To consider other maxima, we rewrite $\vert h(\theta)\vert^2$ as
\begin{equation}
    \vert h(\theta)\vert^2 = \frac{1}{d^2}\left(d+2\sum_{k>\ell}\cos(\theta (\lambda_k-\lambda_\ell))\right),
    \label{eq:app_h2_reformulated}
\end{equation}
which can be obtained by separating the sums into parts with equal and different summation indices.
Now, in order for the right-hand side of \cref{eq:app_h2_reformulated} to be maximized on the integration interval, all cosine terms have to be $1$. A trivial way to achieve this is to make all eigenvalues $\lambda_k$ equal to each other. We exclude this case and ask if there are $\theta\neq 0$ that maximize \cref{eq:app_h2_reformulated}. This requires
\begin{equation}
  \forall k,\ell\in\left[d\right]:\quad\theta(\lambda_k-\lambda_\ell) = 2n\pi\qquad ,n\in\mathbb{Z}.
  \label{eq:cosine_condition}
\end{equation}

Assume a global maximum of \cref{eq:app_h2_reformulated} occurs at $\theta=\pi/a$ for some $a>1$. 
This gives
\begin{equation}
  \forall k,\ell\in\left[d\right]:\qquad \lambda_k-\lambda_\ell=n\frac{\pi}{\theta}=n\cdot a\quad n\in\mathbb{Z}.
  \label{eq:eigenvalue_condition_maxima}
\end{equation}

Since we are dealing with integer eigenvalues the differences must also be an integers, which enforces $a\in\mathbb{N}$.
According to \Cref{eq:eigenvalue_condition_maxima}, the common divisor of all the differences of eigenvalues must be an integer multiple of $a$. We have explained in the main text that the distributions for a set of eigenvalues is related to the distributions for a rescaled set of eigenvalues in a trivial way, see the discussion around \cref{eq:arithmetic_eigenvalues_scaled}. Thus by  shifting first and then rescaling the $\lambda_k$ values, we can reduce the eigenvalues to the case of integer eigenvalues with differences that have the greatest common divisor $1$. Then the only maximum of $|h(\theta)|^2$ occurs at $\theta=0$.

Thus we assume that the only global maximum of $|h(\theta)|^2$ occurs at $\theta=0$. We expand $h(\theta)$ around $0$, which yields
\begin{equation}
    h(\theta)=1+i\theta \bar{\lambda} - \frac{\theta^2}{2}\frac{1}{d}\sum_{k=1}^d \lambda_k^2 + O(\theta^3),
    \label{eq:app_h_expansion}
\end{equation}
with $\bar{\lambda}=\frac{1}{d}\sum_k \lambda_k$ being the average of the eigenvalues.
Again, for small $\theta$ we  have
\begin{equation}
    \log h(\theta) = i\theta\bar{\lambda} - \frac{\theta^2}{2}\frac{1}{d}\sum_{k=1}^d \left(\lambda_k-\bar{\lambda}\right)^2 + O(\theta^3)= i\theta\bar{\lambda} - \frac{\theta^2}{2}s^2 + O(\theta^3),
    \label{eq:app_logh_expansion}
\end{equation}
where 
\begin{equation}
    s^2=\frac{1}{d}\sum_{k=1}^d (\lambda_k-\bar{\lambda})^2
\end{equation}
is the average quadratic deviation of $\lambda_k$ from its mean.
With this we have
\begin{equation}
    \ln \vert h(\theta)\vert^2 = - s^2\theta^2 + O(\theta^4),
    \label{eq:app_logabsh_expansion}
\end{equation}
which results in
\begin{equation}
    \vert h(\theta)\vert^{2L}=\exp\left(-Ls^2 \theta^2+ LO(\theta^4)\right).
    \label{eq:app_h_exponential_correction}
\end{equation}
Now, outside the relevant integration window of $\vert\theta\vert=O(1/\sqrt{L})$ the leading term of \cref{eq:app_h_exponential_correction} is already exponentially suppressed.
The $O(\theta^4)$ contribution in the dominant window at $|\theta|\sim 1/\sqrt{L}$ is of the order
\begin{equation}
    L\cdot O(\theta^4) = L\cdot O\!\left(\frac{1}{L^2}\right) = O\!\left(\frac{1}{L}\right)\xrightarrow{L\to\infty} 0,
    \label{eq:app_correction_asymptotics}
\end{equation}
and can therefore be neglected. Thus in the limit $L\rightarrow\infty$ we have
\begin{equation}
    \vert h(\theta)\vert^{2L}\sim\exp\left(-Ls^2 \theta^2\right).
    \label{eq:app_h_exponential}
\end{equation}
In order to evaluate the integral \cref{eq:app_redundancy_integral}, using the same argument for the correction, we can extend the integration limits from $\pm\pi$ to $\pm\infty$, since the contribution to the integral from $[\pm\pi,\pm\infty)$ is exponentially small compared to the contribution from the $[-\pi,\pi]$ region.
Thus we can write the integral in \cref{eq:app_redundancy_integral} asymptotically as the Gaussian integral
\begin{equation}
    r_\omega \sim \frac{d^{2L}}{2\pi}\int_{-\infty}^{\infty} e^{-i\omega\theta - Ls^2\theta^2}\,d\theta,
    \label{eq:app_approx_integral}
\end{equation}
which evaluates to the expression given in the main text
\begin{equation}
    r_\omega \sim \frac{d^{2L}}{\sqrt{4\pi L s^2}}\exp\!\left(-\frac{\omega^2}{4Ls^2}\right),\qquad L\rightarrow\infty.
    \label{eq:app_approx_integral_solved}
\end{equation}

\twocolumngrid
\bibliographystyle{IEEEtran}
\bibliography{main}
\end{document}